\documentclass[a4paper,onecolumn,11pt,noarxiv]{quantumarticle}

\usepackage{amsmath,amsthm,amsfonts,amssymb,times,bbm,graphicx,mathtools}
\usepackage{xprintlen}
\usepackage[utf8]{inputenc}
\usepackage[english]{babel}
\usepackage[T1]{fontenc}

\usepackage{algorithm}
\usepackage{algorithmicx}
\usepackage{algpseudocode}
\usepackage{xcolor}
\usepackage{hyperref}
\hypersetup{
  colorlinks=true,
  hypertexnames=false,
  linktocpage,
  colorlinks=true, 
  urlcolor=magenta!90!black,    
  linkcolor=blue!60!black, 
  citecolor=black!60 
}
\usepackage[capitalize, compress]{cleveref}
\usepackage[numbers,sort&compress]{natbib}

\usepackage{complexity}
\newclass{\np}{NP}

\newtheorem{theorem}{Theorem}

\newtheorem{lemma}[theorem]{Lemma}
\newtheorem{corollary}[theorem]{Corollary}

\newtheorem{conjecture}[theorem]{Conjecture}

\newcommand{\bra}[1]{\mbox{$\langle #1 |$}}
\newcommand{\ket}[1]{\mbox{$| #1 \rangle$}}

\def\FF{\mathbbm{F}}
\def\NN{\mathbbm{N}}
\def\EE{\mathbbm{E}}

\def\C{\mathcal{C}}

\def\mc{\mathcal}
\def\mb{\mathbb}

\DeclareMathOperator{\rank}{rank}
\DeclareMathOperator{\range}{range}

\DeclareMathOperator{\supp}{supp}

\DeclareMathOperator{\rad}{rad}

\DeclareMathOperator{\Prob}{Prob}

\renewcommand{\vec}{\mathbf}
\newcommand{\mat}[1]{\mathbf{#1}}

\begin{document}

\title{Secret extraction attacks against obfuscated IQP circuits}

\author{David Gross}
\affiliation{Institute for Theoretical Physics, University of Cologne, Germany
}

\email{david.gross@thp.uni-koeln.de}

\author{Dominik Hangleiter}
\affiliation{{Simons Institute for the Theory of Computing, University of California at Berkeley, California, USA}
}
\affiliation{Joint Center for Quantum Information and Computer Science (QuICS), University of Maryland \& NIST, College Park, Maryland, USA
}

\email{mail@dhangleiter.eu}

\date{\today}

\begin{abstract} 
Quantum computing devices can now perform sampling tasks which, according to complexity-theoretic and numerical evidence, are beyond the reach of classical computers.
This raises the question of how one can efficiently verify that a quantum computer operating in this regime works as intended.
In 2008, Shepherd and Bremner proposed a protocol in which a verifier constructs a unitary from the comparatively easy-to-implement family of so-called \emph{IQP circuits}, and challenges a prover to execute it on a quantum computer.
The challenge problem is designed to contain an obfuscated secret, which can be turned into a statistical test that accepts samples from a correct quantum implementation.
It was conjectured that extracting the secret from the challenge problem is \np-hard, so that the ability to pass the test constitutes strong evidence that the prover possesses a quantum device and that it works as claimed.
Unfortunately, about a decade later, Kahanamoku-Meyer found an efficient classical secret extraction attack.
Bremner, Cheng, and Ji very recently followed up by constructing a wide-ranging generalization of the original protocol.
Their \emph{IQP Stabilizer Scheme} has been explicitly designed to circumvent the known weakness.
They also suggested that the original construction can be made secure by adjusting the problem parameters.
In this work, we develop a number of secret extraction attacks which are effective against both new approaches in a wide range of problem parameters.
In particular, we find multiple ways to recover the 300-bit secret hidden in a challenge data set published by Bremner, Cheng, and Ji.
The important problem of finding an efficient and reliable verification protocol for sampling-based proofs of quantum supremacy thus remains open.
\end{abstract}


\maketitle



\section{Introduction}
\label{sec:intro}

A central challenge in the field of quantum advantage is to devise efficient quantum protocols that are both classically intractable and classically verifiable, while minimising the experimental effort required for an implementation. 
The paradigmatic approach satisfying these first conditions is to solve public-key cryptography schemes using Shor's algorithm. 
However, the quantum resources required in the cryptographically secure regime are enormous, using thousands of qubits and millions of gates \cite[see e.g.][]{gidney_how_2019,litinski_how_2023}. 
Reducing the required resources, interactive proofs of computational quantumness have been proposed, which make use of classically or quantum-secure cryptographic primitives \cite{brakerski_cryptographic_2018,brakerski_simpler_2020,kahanamoku-meyer_classically_2022}. 
Again, however, their implementation requires arithmetic operations, putting the advantage regime far beyond the reach of current technology \cite{zhu_interactive_2023}.

A different approach to demonstrations of quantum advantage has focused on simple protocols based on sampling from the output of random quantum circuits \cite{bremner_classical_2010,aaronson_computational_2013,bremner_average-case_2016,boixo_characterizing_2018,arute_quantum_2019_short,zhong_quantum_2020_short}. 
These require a significantly smaller amount of qubits and gates, and seem to be classically intractable even in the presence of noise on existing hardware \cite{wu_strong_2021_short,pan_solving_2022,kalachev_classical_2021,morvan_phase_2023_short}.
However, they are not efficiently verifiable (see Ref.~\cite{hangleiter_computational_2023-1} for a discussion), and already present-day experiments are outside of the regime in which the samples can be efficiently checked.  

The key property which makes random quantum sampling so much more feasible compared to cryptography-based approaches is their apparent lack of structure in the sampled distribution. 
At the same time, this is also what seems to thwart classical verifiability. 
\citet{yamakawa_verifiable_2022} have made significant progress by showing that there are also highly unstructured \np-problems based on random oracles, which can be efficiently solved by a quantum computer and checked by a classical verifier.
Conversely, one may wonder
whether it is possible to introduce just enough structure into random quantum circuits to make their classical outputs efficiently verifiable while keeping the resource requirements low
\cite{shepherd_temporally_2009,aaronson_talk_2022,aaronson_talk_2023}. 
An early and influential idea of this type dating back to 2008 is the one of \citet{shepherd_temporally_2009}. 

\citeauthor*{shepherd_temporally_2009} proposed a sampling-based scheme based on so-called \emph{IQP (Instantaneous Quantum Polynomial-Time)} circuits. 
In the IQP paradigm, one can only execute gates that are diagonal in the $X$-basis. 
They designed a family of IQP circuits based on \emph{quadratic residue codes} (QRC) whose output distribution has high weight on bit strings $\vec x \in \FF_2^n$ that are contained in a hyperplane $\vec s^T \vec x=0 \mod 2$.
The normal vector $\vec s\in\FF_2^n$ may be chosen freely, but its value is not apparent from the circuit description.
This way, a verifier can design a circuit that hides a \emph{secret} $\vec s$.
The verifier then challenges a prover to produce samples such that a significant fraction of them lie in the hyperplane orthogonal to $\vec s$. 
At the time, the only known way to efficiently meet the challenge was for the prover to collect the samples by implementing the circuit on a quantum computer.
More precisely, \citeauthor{shepherd_temporally_2009} conjectured that it was an \np-hard problem to recover the secret from $\mat H$.
They challenged the community to generate samples which have high overlap with a secret 244-bit string---corresponding to a 244-qubit experiment.\footnote{See \href{https://quantumchallenges.wordpress.com}{https://quantumchallenges.wordpress.com} for the challenge of \citet{shepherd_temporally_2009}}

Unfortunately, in 2019, \citet{kahanamoku-meyer_forging_2023} solved the challenge and recovered the secret string. 
The paper provided evidence that the attack has only quadratic runtime for the QRC construction. 

Recently, \citet*{bremner_iqp_2023_v1} have made new progress on this important problem. 
They propose a wide-ranging generalization of the construction---the \emph{IQP Stabilizer Scheme}---which circumvents \citeauthor{kahanamoku-meyer_forging_2023}'s analysis. 
They also conjecture that an associated computational problem---the \emph{Hidden Structured Code (HSC) Problem}---cannot be solved efficiently classically for some parameter choices and pose a challenge for an IQP experiment on 300 qubits, corresponding to a 300-bit secret.
Finally, they also extend the QRC-based construction to parameter regimes in which \citeauthor{kahanamoku-meyer_forging_2023}'s ansatz fails. 

Here, we show that the scheme is still vulnerable to classical cryptanalysis by devising a number of secret extraction attacks against obfuscated IQP circuits.
Our first approach, the \emph{Radical Attack} instantly recovers the 300-bit secret of the challenge from the circuit description. 
We analyse the Radical Attack in detail and give conditions under which we expect the ansatz to work. 
The theory is tested on 100k examples generated by a software package provided as part of \cite{bremner_iqp_2023_v1}, and is found to match the empirical data well.
We also observe that for the Extended QRC construction, the Radical Attack and the approach of \citeauthor{kahanamoku-meyer_forging_2023} complement each other almost perfectly, in the sense that for every parameter choice, exactly one of the two works with near-certainty.

In the final part of the paper, we sketch a collection of further approaches for recovering secretes hidden in IQP circuits.
Concretely, we propose two extensions of \citeauthor*{kahanamoku-meyer_forging_2023}'s idea, which we call the \emph{Lazy Linearity Attack} and the \emph{Double Meyer}. 
The Double Meyer Attack is effective against the Extended QRC construction for all parameter choices, and we expect that its runtime is at most quasipolynomial on all instances of the IQP Stabilizer Scheme. 
Finally, we introduce \emph{Hamming's Razor}, which can be used to identify redundant rows and columns of the matrix that were added as part of the obfuscation procedure. 
For the challenge data set, this allows us to recover the secret in an alternate fashion and we expect it to reduce the load on further attacks in general.

The important problem of finding cryptographic obfuscation schemes for the efficient classical verification of quantum circuit implementations therefore remains open. 

We begin by setting up some notation in \cref{sec:notation}, then recall the IQP Stabilizer Scheme in \cref{sec:construction}, describe and analyze the Radical Attack in \cref{sec:attack}, and sketch further exploits in \cref{sec:further}.

\section{Notation and definitions}
\label{sec:notation}

We mostly follow the notation of Ref.~\cite{bremner_iqp_2023_v1}.
This means using boldface for matrices $\mat M$ and column vectors $\vec v$ (though basis-independent elements of abstract vector spaces are set in lightface).
We write $\mat M_i$ for the $i$-th column of a matrix and $\vec v_i$ for the $i$-th coefficient of a column vector.
By the \emph{support of a set $S \subset \FF_2^m$}, we mean the union of the supports of its elements:
\begin{align*}
	\supp V = \{ i \in [1,m] \,|\, \exists\,\vec v \in V, \>  \vec v_i \neq 0 \}.
\end{align*}
We use $\vec e^i$ for the standard basis vector $(\vec e^i)_j = \delta_{ij}$.
The all-ones vector is $\vec 1$, and for a set $S$, $\vec 1_S$ is the ``indicator function on $S$'', i.e.\ the $i$th coefficient of $\vec 1_S$ is $1$ if $i\in S$ and 0 else.
The \emph{Hamming weight} of a vector $\vec v \in \FF_2^n$ is $|\vec v| \coloneqq \sum_{i \in [n]} \vec v_i$.

\subsection{Symmetric bilinear forms}

Compared to Ref.~\cite{bremner_iqp_2023_v1}, we use slightly more geometric language.
The relevant notions from the theory of symmetric bilinear forms, all standard, are briefly recapitulated here.

Let $V$ be a finite-dimensional vector space over a field $\FF$, endowed with a symmetric bilinear form $\beta(\cdot , \cdot )$.
The \emph{orthogonal complement} of a subset $W\subset V$ is
\begin{align*}
	W^\perp = \{ x\in V \,|\, \beta(x,w) =0 \>\forall\,w\in W\}.
\end{align*}
The \emph{radical} of $V$ is $V \cap V^\perp$, which is the space comprising elements $x\in  V$ such that the linear function $\beta(x, \cdot)$ vanishes identically on $V$.
The space $V$ is \emph{non-degenerate} if $\rad V=\{0\}$.
In this case, for every subspace $W\subset V$, we have that $\dim W^\perp + \dim W = \dim V$.
The subspace $W$ is \emph{isotropic} if $W\subset W^\perp$.
The above dimension formula implies that isotropic subspaces $W$ of a non-degenerate space $V$ satisfy $\dim W \leq \frac12 \dim V$.

Let $\{b^{(i)}\}_{i=1}^k$ be a basis of $V$.
Expanding vectors $x,y\in V$ in the basis,
\begin{align*}
	\beta(x,y)
	=
	\beta\left( 
		\sum_i \vec c_i  b^{(i)},
		\sum_j \vec d_j  b^{(j)},
	\right)
	=
	\sum_{ij} \vec c_i \vec c_j
	\,
	\beta\left( 
		 b^{(i)},
		  b^{(j)} 	
	 \right)
	 =
	 \vec c^T
	 \mat M
	 \vec d,
\end{align*}
where $\vec c, \vec d\in\FF^k$ are column vectors containing the expansion coefficients of $x, y$ respectively, and the \emph{matrix representation} $\mat M$ of $\beta$ has elements
\begin{align*}
	\mat M_{ij}
	=
	\beta\left( 
		 b^{(i)},
		  b^{(j)} 	
	 \right).
\end{align*}
A vector $x\in V$ lies in the radical if and only if its coefficients $\vec c$ lie in the kernel of $\mat M$.

Now assume that $V\subset \FF^m$.
The \emph{standard form} on $\FF^m$ is 
\begin{align*}
	\langle \vec x, \vec y\rangle = 
	\vec x^T \vec y = \sum_{i=1}^m \vec x_i \vec y_i.
\end{align*}
The standard form is \emph{non-degenerate} on $\FF^m$, but will in general be degenerate on subspaces.
Let $\mat H$ be an $m\times k$ matrix
and let $W = \range \mat H \subset V$ be its column span.
The restriction of the standard form to $W$ can then be ``pulled back'' to $\FF^k$ by mapping $\vec c, \vec d \in \FF^k$ to 
\begin{align*}
	\langle \mat H \vec c, \mat H \vec d \rangle
	=
	(\mat H \vec c)^T (\mat H \vec d)
	=
	\vec c^T \big( \mat H^T \mat H \big) \vec d
	=
	\vec c^T  \mat G \vec d,
\end{align*}
where $\mat G = \mat H^T \mat H $ is the \emph{Gram matrix} associated with $\mat H$.
We frequently use the fact that in this context, 
\begin{align}\label{eqn:kernel radical}
	\vec d \in \ker \mat G
	\qquad
	\Leftrightarrow
	\qquad
	\mat H \vec d \in \rad W.
\end{align}

\section{The IQP Stabilizer Scheme}
\label{sec:construction}

\subsection{Hiding a secret string in an IQP circuit}
\label{ssec:hiding}

The IQP Stabilizer Scheme of \citeauthor{shepherd_temporally_2009},
described here following the presentation in Ref.~\cite{bremner_iqp_2023_v1},
uses the tableau representation of a collection of $m$ Pauli matrices on $n$ qubits as an $m \times 2n$ binary matrix.
Since IQP circuits are diagonal in the $X$ basis, we restrict to $X$-type Pauli matrices which are described by $m \times n$ matrices with elements in $\FF_2$.
The tableau matrix $\mat H \in \FF_2^{m \times n}$  determines an IQP Hamiltonian $H = \sum_{i \in [m]} (\prod_{j \in [n]}X_j^{\mat H_{ij}})$, with associated IQP circuit $\omega^{H}$ defined in terms of a phase $\omega$. 
Choosing $\omega = e^{i \pi/4}$ \citeauthor{shepherd_temporally_2009}, observe that the full stabilizer tableau of the state $\omega^H \ket 0$ can be expressed in terms of $\mat H$ and use this fact to find IQP circuits whose output distributions have high weight on a subspace $S_{\vec s} = \{\vec x: \langle \vec x, \mat s \rangle = 0\}$ determined by a secret string $\mat s$.

This is ingeniously achieved as follows.
For $\vec s\in\FF_2^n$, obtain $\mathbf{H}_{\vec s}$ from $\mathbf{H}$ by multiplying its $i$-th row with $(\mathbf{H}\vec s)_i$.
Let $\C_{\vec s} \coloneqq \range  \mathbf{H}_{\vec s}$.
Fix some $g_{\max}\in\NN$.
A vector $\vec s\in\FF_2^n$ is called a \emph{secret} of~$\mat H$ if
\begin{enumerate}
	\item
		the co-dimension of the radical $g \coloneqq \dim \C_{\vec s} - \dim \rad \C_{\vec s}\le g_{\max}$,
		and
	\item
		the radical is doubly even, i.e., for all $\vec x \in \rad \C_{\vec s}$, $|\vec x| = 0\mod 4$.
\end{enumerate}

Given an IQP tableau $\mat H$ with secret $\mat s$, \citeauthor{shepherd_temporally_2009} show that 
\begin{align}
	\Pr_{\vec x \leftarrow D_{\mat H}}[\langle \vec x, \vec s \rangle = 0] = \frac12 (2^{- g/2} + 1),
\end{align}
where $D_{\mat H}(\vec x) = |\bra x \omega^H \ket 0|^2$ is the output distribution of $\omega^H$.
A classical verifier can that efficiently identify samples from the correct distribution by computing their mean overlap with the secret string $\mat s$.

\subsection{Stabilizer construction}
\label{sec:stabconstruction}

We briefly recap the specifics of how \citet*{bremner_iqp_2023_v1} construct a pair $(\mat H, \vec s)$, comprising a generator matrix $\mat H$ and corresponding secret $\vec s$.
Pre-obfuscation, the matrix $\mat H$ is of the following form:
\begin{align}\label{eqn:blockmat}
	\vcenter{\hbox{\includegraphics{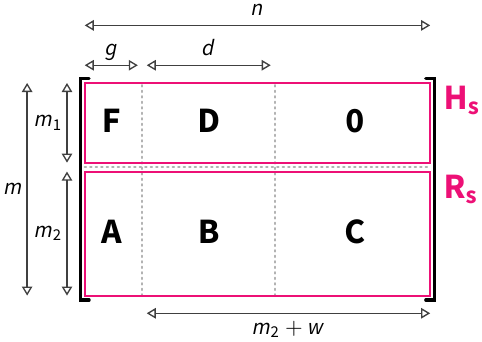}}}
\end{align}
Essentially, the blocks are chosen uniformly at random, subject to the following constraints:
\begin{enumerate}
	\item
		$\mat D$ is an $m_1 \times d$-matrix.
		Its range is a $d$-dimensional doubly-even isotropic subspace of $\FF_2^{m_1}$.
		These constraints are equivalent to
		\begin{align}\label{eqn:condD}
			|\mat D_i| = 0 \mod 4 \>  (i = 1, \dots, d),
			\qquad
			\mat D^T \mat D = 0, \qquad \rank \mat D = d.
		\end{align}
	\item
		$\mat F$ is an $m_1 \times g$-matrix.
		It generates a $g$-dimensional space that is non-degenerate and orthogonal to $\range \mat D$ with respect to the standard form on $\FF_2^{m_1}$.
		These constraints are equivalent to
		\begin{align}\label{eqn:condF}
			\rank\mat F^T\mat F = g,
			\qquad
			\mat D^T \mat F = 0.
		\end{align}
	\item
		There exists a \emph{secret} $\vec s \in \FF_2^n$ such that the inner product between $\vec s$ and the rows of $\mat H$ is non-zero exactly for the first $m_1$ rows:
		\begin{align}\label{eqn:secret}
			\mat H \vec s = \vec 1_{[1,\dots,m_1]}.
		\end{align}
	\item
		Finally, $\mat R_{\vec s} = (\mat A\,|\,\mat B\,|\,\mat C)$ are ``redundant rows'', chosen such that $\mat R_{\vec s} \vec s = \vec 0$ and
		\begin{align}\label{eqn:condR}
			\rank \mat H = n 
			.
		\end{align}
\end{enumerate}
Further comments:
\begin{itemize}
	\item
	Introducing notation not used in \cite{bremner_iqp_2023_v1}, we split $\mat R_{\vec s}$ into sub-matrices $\mat R_{\vec s} = (\mat A\,|\,\mat B\,|\,\mat C)$ according to $\FF_2^n = \FF_2^g \oplus \FF_2^d \oplus \FF_2^{n-g-d}$.
\item
	It will turn out that the parameter
	\begin{align*}
		w:=n-g-m_2
	\end{align*}
	plays a central role for the performance the Radical Attack.
	It may be described as measuring 
	the degree to which the matrix $(\mat B\,|\,\mat C)$ is ``wide'' rather than ``tall''.
	\item
		The range of $\mat H_{\vec s}$ is the code space $\C_{\vec s}$.
		The range of $\mat D$ is its radical $\rad \C_{\vec s}\subset \C_{\vec s}$.
		The range of $\mat F$ is a 
		subspace that is complementary to the radical within the code space.
	\item
		The rank constraint (\ref{eqn:condR}) implies
	\begin{align}\label{eqn:c-has-full-rank}
		\rank \mat C = n - g - d.
	\end{align}
	\item
		There are some subtleties connected to the way the redundant rows are generated according to the paper \cite{bremner_iqp_2023_v1} and the various versions of the software implementation provided.
		See \cref{sec:implementation-details} for more details.
\end{itemize}

The parameters 
$n$ (number of qubits),
$m$ (terms in the Hamiltonian),
and $g$ (log of the power of the statistical test)
are supplied by the user, while $m_1$ and $d$ are chosen randomly.
The precise way in which $m_1, d$ are to be generated does not seem to be specified in the paper, so we take guidance from the reference implementation provided at Ref.~\cite{cheng_repo}.
Their \verb|sample_parameters()|-function (found in \verb|lib/construction.py|) fixes these numbers in a two-step procedure.
First, preliminary values of $m_1/2$ and $d$ are sampled according to binomial distributions with parameters roughly given as 
\begin{align}\label{eqn:binom}
	m_1 /2 \sim \operatorname{Bin}\left( N \approx \frac{m-g}2, p=0.3\right),
	\qquad
	d \sim \operatorname{Bin}\left( N = \left\lfloor\frac{m_1-g}2\right\rfloor, p=0.75\right).
\end{align}
The values are accepted if they satisfy the constraints
\begin{align}\label{eqn:constraints}
	n-g
	\leq
	d
	\leq
	w.
\end{align}
We are not aware of a simple description of the distribution conditioned on the values passing the test.
Empirically, we find that for the challenge parameters
\begin{align}\label{eqn:challenge}
	n = 300,
	\quad
	m = 360,
	\quad
	g = 4,
\end{align}
the following values are attained most frequently:
\begin{align}\label{eqn:reference}
	m_1 = 102, 
	\quad
	d = 38
	\qquad
	\Rightarrow
	\qquad
	m_2 = 258, 
	\quad
	w = 38.
\end{align}

Given $(\mat H, \mat s)$, obfuscation is then performed as $\mat H \leftarrow \mat P \mat H \mat Q$, $\mat s \leftarrow \mat Q^{-1} \mat s$ using a random  invertible matrix $\mat Q$ and a random (row-)permutation $\mat P$.  
\smallskip

\citeauthor*{bremner_iqp_2023_v1} pose the following conjecture: 
\begin{conjecture}[Hidden Structured Code (HSC) Problem \cite{bremner_iqp_2023_v1}]
\label{conj:hsc}
	For certain appropriate choices of $n, m, g$, there exists an efficiently samplable distribution over instances $(\mat H, \mat s)$ from the family $\mc H_{n,m,g}$, so that no polynomial-time classical algorithm can find the secret $s$ given $n$, $m$ and $\mat H$ as input, with high probability over the distribution on $\mc H_{n,m,g}$.
\end{conjecture}

\subsection{Extended QRC construction}
\label{sec:extended qrc}

In the original QRC construction of \citet*{shepherd_temporally_2009}, $(\mat F\,|\,\mat D)$ is chosen as a $q \times (q+1)/2$ quadratic residue code with prime $q$ such that $q + 1 \mod 8 = 0$.
Then, the all-ones vector $\vec 1_q$, which is guaranteed to be a codeword of a QRC, is appended as the first column of $(\mat F \, | \, \mat D)$. 
Next, rows are added which are uniformly random, except for the first entry which is $0$. 
This ensures that there is a secret $\vec s = \vec e_1$.
The resulting matrix is then obfuscated as above. 

In the Extended QRC construction, additional redundant columns are added, essentially amounting to a nontrivial choice of $\mat C$, in order to render the algorithm of \citeauthor*{kahanamoku-meyer_forging_2023} ineffective.
Letting $r = (q+1)/2$, \citeauthor*{bremner_iqp_2023_v1} propose to add $q$ redundant rows to achieve $m = 2 q $, and add a redundant $\mat C$ block to achieve a width $n$ satisfying $r \leq n \leq  q + r $. 

\section{The Radical Attack}
\label{sec:attack}

Starting point of the attack was the empirical observation that $\mat H^T \mat H$ for the challenge generator matrix has a much larger kernel ($\dim\ker\mat H^T \mat H = 34$) than would be expected for a random matrix of the same shape 
(about $1$).
This observation gave rise to the Radical Attack, summarized in \cref{alg:radicalattack}. 

\begin{algorithm}
	\caption{Radical Attack}\label{alg:radicalattack}
	\begin{algorithmic}[1]
		\Function{RadicalAttack}{$\mat H$}
	\State $\mat G\leftarrow \mat H^T \mat H$
	\State $\mat K\leftarrow $ a column-generating matrix for $\ker \mat G$
	\State $S\leftarrow $ the support of the columns in $\mat H\mat K$
	\State Solve the $\FF_2$-linear system $\mat H \vec s = \vec 1_S$
	\State \Return $\vec s$
	\EndFunction
	\end{algorithmic}
\end{algorithm}

We have tested this ansatz against instances created by the software package provided by \citeauthor*{bremner_iqp_2023_v1}.
For the parameters $(n,m,g)=(300,360,4)$, used for the challenge data set, the secret is recovered with probability about $99.85\%$.
The challenge secret itself can be found using a mildly strengthened version.

We will analyze this behavior theoretically in Sec.~\ref{sec:analysis}, report on the numerical findings in Sec.~\ref{sec:numerics}, and, in Sec.~\ref{sec:challenge}, explain why the challenge Hamiltonian requires a modified approach.

\subsection{Performance of the attack}
\label{sec:analysis}

The analysis combines three ingredients:
\begin{enumerate}
	\item
		We will show that with high probability, $\mat H(\ker \mat G)$ is a subspace of the radical $\rad\C_{\vec s}$.
		Because $\mat H$ and thus $\mat G =\mat H^T \mat H$ are known, this means that we can computationally efficiently access elements of the radical.
	\item
		We will then show that the intersection of $\mat H(\ker \mat G)$ with the radical is expected to be relatively large.
\end{enumerate}
These two statements follow as \cref{cor:kernel} from a structure-preserving normal form for obfuscated generator matrices of the form \eqref{eqn:blockmat}, described in \cref{lem:normalform}.
\begin{enumerate}
	\setcounter{enumi}{2}
	\item
		In \cref{lem:support}, 
		we argue that one can expect that the non-zero coordinates that show up in this subspace coincide with the obfuscated coordinates $1 \ldots m_1$.
\end{enumerate}

\subsubsection{A normal form for generator matrices}

Recall the notion of \emph{elementary column operations} on a matrix, as used in the context of Gaussian elimination.
Over $\FF_2$, these are (1) exchanging two columns, and (2) adding one column to another one.
Performing a sequence of column operations is equivalent to applying an invertible matrix from the right.
We will map the generator matrix $\mat H$ to a normal form using a restricted set of column operations. These column operations preserve the properties of the blocks of $\mat H$  described in \cref{sec:construction}.

To introduce the normal form, split $\mat H$ into blocks as
\begin{align*}
	\mat H
	=
	(\mat{\hat A}\,|\,\mat{\hat B}\,|\,\mat{\hat C}),
	\qquad
	\hat{\mat A} := \begin{pmatrix} \mat F \\ \mat A \end{pmatrix},
	\quad
	\hat{\mat B} := \begin{pmatrix} \mat D \\ \mat B \end{pmatrix},
	\quad
	\hat{\mat C} := \begin{pmatrix} \mat 0 \\ \mat C \end{pmatrix}.
\end{align*}
We say that an elementary column operation \emph{is directed to the left} if it
\begin{itemize}
	\item
		adds a columns of $\hat{\mat C}$ to another colum in $(\hat{\mat A}\,|\,\hat{\mat B}\,|\,\hat{\mat C})$, 
	\item
		adds a column of $\hat{\mat B}$ to another column in $(\hat{\mat A}\,|\,\hat{\mat B})$, 
	\item
		adds a column of $\hat{\mat A}$ to another column in $\hat{\mat A}$, or
	\item
		permutes two columns within one block.
\end{itemize}

The first part of the following lemma lists properties that are preserved under such column operations.
The second part describes two essential simplifications to $\mat H$ which can still be achieved.

\begin{lemma}[Normal form]
	\label{lem:normalform}
	Assume $\mat H'$ results from $\mat H$ by a sequence of column operations that are directed to the left.
	Then:
	\begin{enumerate}
		\item
			If $\mat H$ is a block matrix of the form (\ref{eqn:blockmat}) and fulfills the conditions (\ref{eqn:condF}) -- (\ref{eqn:condR}), then the same is true for $\mat H'$. 
		\item
			It holds that:
			\begin{align*}
				\range (\mat F'\,|\,\mat D') &= \range (\mat F\,|\,\mat D),  \\
				\range \mat D' &= \range \mat D,  \\
				\range (\mat B'\,|\,\mat C') &= \range (\mat B\,|\,\mat C)  .
			\end{align*}
	\end{enumerate}

	There is a sequence of column operations directed to the left such that:
	\begin{enumerate}
	\setcounter{enumi}{2}
		\item
			If $\range(\mat B\,|\,\mat C) = \FF_2^{m_2}$, then $\mat A'=0$.
		\item
			In any case,
			$\dim\ker\mat B' \geq w $. 
	\end{enumerate}
\end{lemma}

\begin{proof}
	Claim~1 and 2 follow directly from the definitions.
	Least trivial is the statement that Condition~(\ref{eqn:condF}) is preserved, so we make this one explicit.
	Consider the case where the $i$-th column of $\hat{\mat B}$ is added to the $j$-th column of $\hat{\mat A}$.
	This will change $\mat F \mapsto \mat F'=\mat F + \mat D_i\, (\vec e^j)^T$.
	But then, by (\ref{eqn:condD}),
	\begin{align*}
		(\mat F')^T \mat F'
		=
		\mat F^T \mat F 
		+ \vec e^j \, \mat D_i^T \mat F
		+ \mat F^T \mat D_i\,(\vec e^j)^T
		+ \vec e^j\,\mat D_i^T \mat D_i\,(\vec e^j)^T
		=
		\mat F^T \mat F.
	\end{align*}

	Claim~3 is now immediate:
	Assuming the range condition, every column of $\mat A$ can be expressed as a linear combination of columns in $(\mat B\,|\,\mat C)$. 
	Therefore, $\mat A$ may be eliminated by column operations directed to the left.

	To prove Claim~4, choose a basis $\{\vec b^i\}_{i=1}^k$ for $\range\mat B \cap \range\mat C$.
	Using column operations within $\hat{\mat B}$ and $\hat{\mat C}$ respectively, we can achieve that the first $k$ columns of $\mat B$ and of $\mat C$ are equal to $\vec b^1, \ldots, \vec b^k$.
	The first $k$ columns of $\mat B$ can then be set to zero by subtracting the corresponding columns of $\mat C$.
	Using \cref{eqn:c-has-full-rank} and the trivial bound on $\dim(\range \mat B \cap \range \mat C)$,
	\begin{align*}
		k=
		\dim (\range \mat B \cap \range \mat C) 
		\geq
		\rank \mat B + \rank \mat C - m_2  
		=
		(\rank \mat B -d) + 
		n - g - m_2 
	\end{align*}
	so that the kernel of the resulting matrix $\mat B'$ satisfies 
	\begin{align*}
		\dim\ker\mat B'
		=
		d-\rank \mat B + k
		\geq	
		n - g - m_2 
		=w.
	\end{align*}
\end{proof}

\subsubsection{Accessing elements from the radical}

As alluded to at the beginning of this section, the normal form implies that $\mat H(\ker \mat G)$ is expected to be a subspace of $\rad\C_{\vec s}$, which is fairly large.
More precisely:

\begin{corollary}
	\label{cor:kernel}
	We have that:
	\begin{enumerate}
		\item
			If
			$\range (\mat B\,|\,\mat C) = \FF_2^{m_2}$,
			then  
			$\mat H(\ker \mat G)\subset 
			\big((\rad \C_{\vec s}) \oplus 0\big)$.

		\item
			In any case,
			$\dim\left(\rad \C_{\vec s} \cap \mat H (\ker \mat G) \right)
			\geq
			w$.

		\end{enumerate}
\end{corollary}

\begin{proof}
	By \cref{lem:normalform} and \cref{eqn:kernel radical}, the advertised statements are invariant under column operations directed to the left.

	If
	$\range (\mat B\,|\,\mat C) = \FF_2^{m_2}$,
	we may thus assume that $\mat A=0$, which gives
	\begin{align*}
		\mat G
		&=
		\begin{pmatrix}
			\mat F^T \mat F & 0 							& 0 \\
			0					 			& \mat B^T \mat B & \mat B^T \mat C \\
			0					 			& \mat C^T \mat B & \mat C^T \mat C
		\end{pmatrix}.
	\end{align*}
	Because $\mat F^T \mat F$  has full rank, $\ker \mat G = 0 \oplus (\ker (\mat B\,|\,\mat C))$.
	But elements of this space are mapped into $\rad\C_{\vec s}$ under $\mat H$.
	This proves the first claim.

	From the block form~\eqref{eqn:blockmat} and the fact that 
	$\mat D$ 
	is non-degenerate, it follows that $\mat H$ embeds $0\oplus(\ker \mat B)\oplus 0  \subset \ker \mat G$ into $\rad\C_{\vec s}$.
	Claim 2 then follows from $\dim\ker\mat B \geq n - g - m_2$, which we may assume since the claim is invariant under column operations directed to the left.
\end{proof}

If we model $\mat B, \mat C$ as random matrices with elements drawn uniformly from $\FF_2$, the probability that $\range (\mat B\,|\, \mat C) = \FF_2^{m_2}$ can be estimated from the well-studied theory of random binary matrices.
Indeed, in the limit $k\to\infty$, 
the probability $\rho(w)$ that a random binary $k\times (k+w)$ matrix has rank less than $k$ is given by
\begin{align*}
	\rho(w)
	=	
	1-\prod_{i=w+1}^\infty \left(1-2^{-i}\right),
\end{align*}
c.f.~\cite[Thm~3.2.1]{kolchin_random_1998}.
This expression satisfies
\begin{align}\label{eqn:pochhammer approx}
	2^{-w} \leq \rho(w) \leq 2^{-w+1},
	\qquad
	\lim_{w\to\infty} \rho(w) 2^{w} = 1,
\end{align}
and one may verify on a computer that $2^{-w}$ is an excellent multiplicative approximation to $\rho(w)$ already for $w\approx 7$.
Thus, interestingly, the value of $w$ governs the behavior of both parts of Corollary~\ref{cor:kernel}.

\subsubsection{Reconstructing the support from random samples}

We proceed to the third ingredient of the analysis---asking whether the support of the numerically obtained elements of the radical is likely to be equal to the obfuscated first $m_1$ coordinates.

\begin{lemma}
\label{lem:support}
	Let $V$ be a subspace of $\FF_2^{m_1}$.
	Take $k$ elements $\{\vec v^{i}\}_{i=1}^k$ from $V$ uniformly at random.
	The probability that $\supp\left(\{ \vec v^{i} \}_{i=1}^k\right) \neq \supp V$ is no larger than $m_1 2^{-k}$
\end{lemma}

\begin{proof}
	Let $j\in\supp V$. 
	We can find a basis $\vec b^{j}$ of $V$ such that exactly $\vec b^{1}$ is non-zero on the $j$-th coordinate.
	Therefore, for each $j$, exactly half the elements of $V$ of are non-zero on $j$.
	Thus, the probability that $j$ is not contained in the support of the vectors is $2^{-k}$.
	The claim follows from the union bound.
\end{proof}

To apply the lemma to the situation at hand, let us again adopt a simple model where the blocks of $\mat H$ are represented by uniformly random matrices.
Under this model, we expect 
$\supp\range\mat D = [m_1]$ 
to hold if $d>\log_2 m_1$, and, in turn, 
$\supp\mat H(\ker \mat G) = \supp\range\mat D$
if $w > \log_2 m_1$.
Again, the probability of failure decreases exponentially in the amount by which these bounds are exceeded.
For the reference values (\ref{eqn:reference}), $\log_2 m_1 \approx 6.67$. 

While it is highly plausible that the uniform random model accurately captures the distribution of $\mat H(\ker \mat G)$,
this is less obvious for $\mat D$, which is constrained to have doubly-even, orthogonal and linearly independent columns. 
Nonetheless, it will turn out that the predictions made based on this model fit the empirical findings very well.
This suggests that in the choice of $\mat D$, full support on~$[m_1]$ is attained at least as fast as suggested by the random model. 
We leave finding a theoretical justification for this behavior as an open question.\footnote{
	One difficulty to overcome is that it is not apparent that sequential sampling algorithms, like the one implemented in \cite{bremner_iqp_2023_v1}, produce a uniform distribution over all subspaces compatible with the geometric constraints.
	Usually, such results are proven by invoking a suitable version of Witt's Lemma to establish that any partial generator matrix can be extended to a full one in the same number of ways.
	In geometries that take Hamming weight modulo $4$ into account, there may however be obstructions against such extensions.
	A relevant reference is \cite[Sec.~4]{wood_witt_1993}
	(see also Refs.~\cite{gross_schurweyl_2021, montealegre_duality_2022} for a discussion in the context of stabilizer theory).
	The theorem in that section states that two isometric subspaces can be mapped
	onto each other by a global isometry only if they both contain the all-ones
	vector or if neither of them does.
	The resulting complications have prevented us from finding a simple rigorous version of \cref{lem:support} that applies to random generator matrices of doubly-even spaces.
}

The factor $m_1$ in the probability estimate of \cref{lem:support} comes from a union bound, and is tight only in the unrealistic case where for every possibly choice of $\{\vec v^i\}_{i=1}^k$, at most one element is missing from the support.
A less rigorous, but plausibly more realistic, estimate of the error probability is obtained if we assume that the coefficients $\vec v_j$ of random elements $\vec v$ of $V$ are distributed independently.
In this case, the error probability is $(1-2^{-k})^{m_1}$ rather than $m_1 2^{-k}$.

\subsubsection{Global analysis}

Combining the various ingredients, we can estimate the probability of recovering the secret given~$w$.
If all conditions are modelled independently (rather than using more conservative union bounds), the result is
\begin{align}\label{eqn:independent success rate}
	\Prob[ \text{success} \,|\, w]
	\approx
	(1-\rho(w))
	(1-2^{-w})^{m_1}
	(1-2^{-d})^{m_1}.
\end{align}

A number of simplifying approximations are possible.
The approximate validity of some of these steps, like dropping the ``$+1$'' in the exponent in the following displayed equation, are best verified by graphing the respective curves on a computer.

From Eq.~(\ref{eqn:pochhammer approx}),
\begin{align*}
	\Prob[ \text{success} \,|\, w]
	\approx
	(1-2^{-w})^{m_1+1}
	(1-2^{-d})^{m_1}
	\approx
	(1-2^{-w})^{m_1}
	(1-2^{-d})^{m_1}.
\end{align*}
Next, we argue that the dependency on $d$ can be neglected.
Indeed, the constraints (\ref{eqn:constraints}) enforce $d\geq w$, so that the success probability differs significantly from $1$ if and only if $w$ is small. 
Now recall from Eq.~(\ref{eqn:binom}) that, conditioned on $w$, the value of $d$ is
sampled from a binomial distribution with expectation value
\begin{align*}
	\EE[ d \,|\, w ]
	\approx
	\frac34
	\frac{m_1-g}{2}
	=
	\frac38w
	+
	\frac38
	( g + m - n)
\end{align*}
and then postselected to satisfy $g-n \leq d \leq w$.

For $(n,m,g)=(300,360,4)$ and $w\leq 20$, the expectation value 
$\EE[ d \,|\, w ] \approx 0.375 w + 20$ 
is sufficiently far away from the boundaries 
imposed by the constraint that the effects of the post-selection may be neglected.
Then, in this parameter regime, we expect $d\simeq w + 20$, so that $2^{-d} \ll 2^{-w}$.

Hence
\begin{align*}
	\Prob[ \text{success} \,|\, w]
	&\approx
	(1-2^{-w})^{m_1}
	=
	(1-2^{-w})^{w+g+m-n}
	\approx
	(1-2^{-w})^{g+m-n}.
\end{align*}
The final expression is a sigmoid function which reaches the value $1/2$ at
\begin{align*}
	w_{\frac12} = - \log_2\left(1-2^{\frac1{g+m-n}}\right)\approx 6.3,
\end{align*}
and we therefore predict that the probability of success of the Radical Attack transitions from~$0$ to~$1$ around a value of~$w$ between 6 and 7.

\subsection{Numerical experiments for the stabilizer construction}
\label{sec:numerics}

\begin{figure}
\centering
\includegraphics{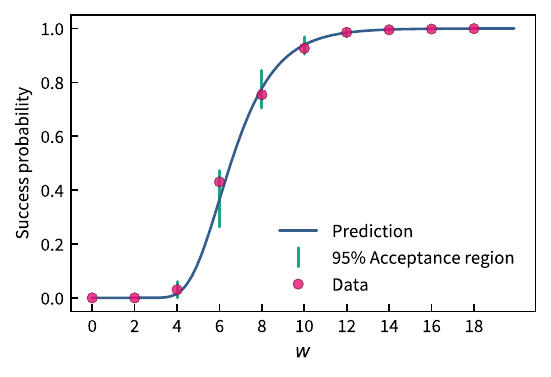}
	\caption{\label{fig:sigmoid} 
		Probability of success of the Radical Attack given the ``excess width'' $w$ of the matrix $(\mat B\,|\,\mat C)$.
		The solid sigmoidal curve is the simplified theoretical estimate $\Prob[\text{success}\,|\,w]\approx (1-2^{-w})^{g+m-n}$.
		Red dots represent empirical success probabilities for all values of $w$ for which failures have been observed during 100k numerical runs.
		Each vertical bar is the acceptance region of a test for compatibility with the theory prediction at significance level $\alpha=5\%$.
		The plot is truncated at $w=18$, as this is the largest value for which {\sc RadicalAttack}$()$ has failed at least once to recover the correct secret in the experiment.
		The mean value of $w$ is about $32.3$ and more than 96\% of all instances were associated with a value of $w$ exceeding 18.
		The simplified theoretical analysis reproduces the behavior of the algorithm in a quantitatively correct way, including predicting the transition from likely failure to likely success at around $w\approx 6$.
		}
\end{figure}

We have sampled $\gtrapprox$ 100k instances of $\mat H$ for $m=360, n=300, g=4$; see \cite{repository} for computer code and the raw data.
Only in 154 cases did the Radical Attack fail to uncover the secret.

Given the number of approximations made, the theoretical analysis turns out to give a surprisingly accurate quantitative account of the behavior of the attack.
This is visualized in Fig.~\ref{fig:sigmoid}.
In particular, the transition from expected failure to expected success around $w_{\frac12}\simeq 6.3$ can be clearly seen in the data.
Consistent with the theory, no failures were observed for instances with $w>18$.

\subsubsection{Uncertainty quantification for the numerical experiments}

Because many of the predicted probabilities are close to $0$ or $1$, finding a suitable method of uncertainty quantification is not completely trivial.

Commonly, when empirical findings in the sciences are compared to theoretical predictions, one computes a confidence interval with coverage probability $(1-\alpha)$ for the estimated quantity and checks whether the theory prediction lies within that interval.
Operationally, this furnishes a statistical hypothesis test for the compatibility between data and theory at significance level $\alpha$ 
(i.e.\ the probability that this method will reject data that is in fact compatible is at most $\alpha$).
However, among the set of all hypothesis tests at a given significance level, some are more powerful than others, in the sense that they reject more data sets.
The common method just sketched turns out to be of particularly low power in our setting.

Indeed, consider the extreme case where the hypothesis is $X \sim \operatorname{Binom}(N,p=0)$.
Then a single instance of a non-zero outcome $X_i=1$ is enough to refute the hypothesis at any significance level. 
In other words, the acceptance region for the empirical probability $\hat p = |\{ i\,|\, X_i=1\}|/N$ is just $\{0\}$.
On the other hand, the statistical \emph{rule of three} states that if no successes have been observed in $N$ attempts, a $95\%$-confidence region needs to have size about $3/N$.

Happily, for testing compatibility with the predicted parameter of a binomial distribution, there is a \emph{uniformly most powerful unbiased} (UMPU) test \cite[Chapter~6.2]{shao_mathematical_2003}.
The vertical bars in Fig.~\ref{fig:sigmoid} represent the resulting acceptance region.
We re-iterate that this test is much more stringent than the more common approach based on confidence intervals would be.

\subsection{The challenge data set}
\label{sec:challenge}

In light of the very high success rate observed on randomly drawn examples, it came as a surprise to us that an initial version of our attack \emph{failed} for the challenge data set that the authors provided in Ref.~\cite{cheng_repo}.
Fortunately, \citeauthor*{bremner_iqp_2023_v1} were kind enough to publish the full version control history of their code \cite{cheng_repo}.
The challenge was added with commit \texttt{d485f9}.
Later, commit \texttt{930fc0} introduced a bug fix in the row redundancy routine.
Under the earlier version, there was a high probability of the $\range(\mat B\,|\,\mat C)=\FF_2^{m_2}$-condition failing.
In this case, elements of $\ker \mat G$ would not necessarily correspond to elements of the radical.
However, in the challenge data set, the doubly even part of $\mat H(\ker\mat G)$ \emph{is} contained in the radical.
A minimalist fix---removing all singly-even columns from the generator matrix for $\mat H (\ker \mat G)$---suffices to recover the hidden parameters:
\begin{align*}
	g = 4, 
	\quad
	d = 35,
	\quad 
	m_1 = 96,
\end{align*}
and the secret
\begin{center}
	$\vec s=$
	\verb|cyCxfXKxLxXu3YWND2fSzf+YKtZJFLWY1J0l2rBao0A5zVWRSKA=|
\end{center}
given here as a base64-encoded binary number.
The string has 
since been kindly confirmed as being equal to the original secret by Michael Bremner, Bin Cheng, and Zhengfeng Ji.

\subsection{Application to the Extended Quadratic Residue Code construction}

The Radical Attack performs even better against the QRC construction with parameters
\begin{align*}
	q \in \{ 103,127, 151, 167,223\},
	\quad
	r = \frac{q+1}2, 
	\quad
	m = 2 q,
	\quad
	n = r +q
\end{align*}
recommended by \citeauthor*{bremner_iqp_2023_v1} as most resilient against the Kahanamoku-Meyer approach, see \cref{sec:extended qrc}.
In 20k runs we have found not a single instance in which the Radical Attack fails for these parameter choices; see Ref.~\cite{repository} for code and raw data. 

\begin{figure}
\centering
	\includegraphics{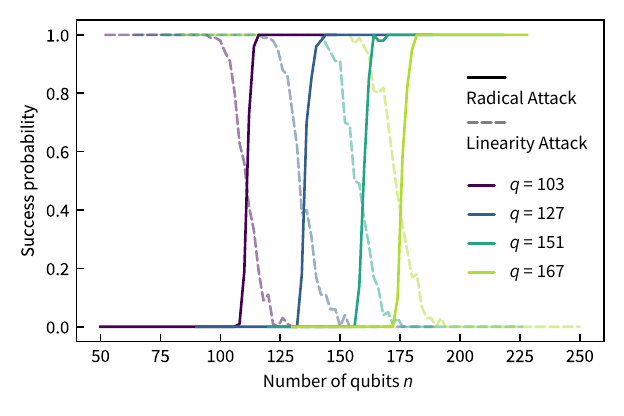}
	\caption{\label{fig:meyer vs radical} 
		Performance of the Radical Attack and the Linearity Attack \cite{kahanamoku-meyer_forging_2023} on the updated QRC construction of \citet*{bremner_iqp_2023_v1}, using 100 random instances per point. 
		Linearity Attack data from \cite[Fig.~3b]{bremner_iqp_2023_v1}. 
		The two approaches are seen to complement each other almost perfectly.
	}
\end{figure}

The Extended QRC construction does not fix $n$ to $r+q$, but rather allows for all values between $r$ and $q + r$. 
This raises the possibility that there is a parameter regime in which both the Linearity Attack and the Radical Attack fail. 
In \cref{fig:meyer vs radical} we explore this possibility. 
We find that the Radical Attack succeeds with high probability for $n \gtrsim q + 13$. 
As discussed after \cref{cor:kernel}, this matches the regime in which we expect
\begin{enumerate}
	\item 
		the rank of the added row redundancy to saturate such that the condition $\range(\mat B | \mat C) = \FF_2^{m_2}$ of \cref{cor:kernel} is satisfied, and 
	\item 
		the parameter $w = n  - q -1$ to exceed $\log_2 m_1 = \log_2 q $, which for the choices of $q = 103, 127, 151, 167$ is given by $\log_2 q \approx 7$. 
\end{enumerate}
Let us also note that the QRC code is guaranteed to have the $\mat 1_{[m_1]} = \vec 1 _{[q]}$ vector as a codeword and hence, it is guaranteed to have full support on the obfuscated coordinates. 

Comparing this performance with the Linearity Attack, we find only a very slim region around $n \approx q + 13$ in which there exist instances that cannot be solved by either attack with near-certainty. 
This motivates the exploration of further cryptanalytic approaches in 
\cref{sec:further}, where we will indeed present two algorithms---the \emph{Lazy Linearity Attack} and the \emph{Double Meyer}, both building on the appproach of \citet{kahanamoku-meyer_forging_2023}---that will eliminate the remaining gap.

\section{Further Attacks}
\label{sec:further}

The Stabilizer IQP Scheme features a large number of degrees of freedom that may allow an algorithm designer to evade any given exploit.
The purpose of this section is to exhibit a variety of further approaches that might aid the cryptanalysis of obfuscated IQP circuits.
Because the goal is to give an attacker a wide set of tools that may be adapted to any particular future construction, we focus on breadth and, as compared to Sec.~\ref{sec:attack}, put less emphasis on rigorous arguments.

\subsection{The Lazy Linearity Attack}
\label{sec:lazy linearity attack}

We begin by slightly extending the Linearity Attack to what we call the \emph{Lazy Linearity Attack}, summarized in \cref{alg:lazymeyer}. 
In addition to the IQP tableau $\mat H$, this routine requires additional input parameters which we call the \emph{ambition} $A$, the \emph{endurance} $E$, and the \emph{significance threshold}~$g_{\text{th}}$.


\begin{algorithm}
	\caption{Lazy Linearity Attack}\label{alg:lazymeyer}
	\begin{algorithmic}[1]
	\Function{LazyLinearityAttack}{$\mat H, g_{\text{th}}, A, E$}
	\While {$\epsilon < E $}
		\State Draw a uniformly random $\vec d \leftarrow \FF_2^n$
		\State $\mat G_{\vec d} \leftarrow \mat H_{\vec d}^T \mat H_{\vec d}$
		\If {$\dim \ker \mat G_{\vec d} < A$}
			\For{$\vec x \in \ker \mat G_{\vec d}$}
			\If{$\rad \range( \vec H_{\vec x}) \neq \{0\}$ and doubly-even and $\rank (\mat H_{\vec x}^T\mat H_{\vec x}) \le g_{\text{th}}$}
			\State \Return $\vec x$ and \textbf{exit}.
			\EndIf 
			\EndFor
			
		\EndIf	
		\State $\epsilon \leftarrow \epsilon +1$
	\EndWhile
	\State \Return ``fail''
	\EndFunction
	\end{algorithmic}
\end{algorithm}

\subsubsection{Analysis}
\label{sec:analysis lazy}

We start by briefly recapitulating why the Linearity Attack of \citet{kahanamoku-meyer_forging_2023} is effective. 
Essentially, it is based on the following property of the kernel of the Gram matrix $\mat G_{\vec d}$ for vectors $ \vec d \in \FF_2^n$. 

\begin{lemma}
 	For $\vec d \in \FF_2^n$, let $\mat G_{\vec d} \coloneqq \mat H_{\vec d}^T \mat H_{\vec d}$. The following implication is true 
 	\begin{align}
 		\mat H_{\vec s}\vec d\in\rad\C_{\vec s} \quad \Rightarrow \quad\vec s \in \ker(\mat G_{\vec d}). 
 	\end{align}
\end{lemma} 
\begin{proof}
By definition, it holds that $\mat H_{\vec s}\vec d\in\rad\C_{\vec s}$ iff
\begin{align}
	\vec d^T \mat H_{\vec s}^T \mat H_{\vec s} \vec e 
	=
	\vec d^T \mat H^T \mat H_{\vec s} \vec e 
	= 0 \qquad \forall\,\vec e \in \FF_2^n.
\end{align}
Because $\vec s \mapsto \mat H_{\vec s}$ is linear, the above means that every element $\mat H {\vec d}\in\rad\C_{\vec s}$ of the radical gives rise to a set of linear equations (one for each $\vec e\in\FF_2^n$) for the secret $\vec s$.
These equations can be compactly written as
\begin{align}
	\vec d^T \mat H^T \mat H_{\vec s}  = 0
	\quad\Leftrightarrow\quad
	\mat H^T \mat H_{\vec d} \vec s  = 0
	\quad\Leftrightarrow\quad
	\mat H_{\vec d}^T \mat H_{\vec d} \vec s  = 0.
\end{align}
 \end{proof}

In the Linearity Attack, the strategy is now to pick $\vec d$ at random.
Then with probability 
\begin{align}
	\frac{|\rad \C_{\vec s}|}{|\C_{\vec s}|}
	=
	\frac{2^{\dim \rad \C_{\vec s}}}{2^{\dim \C_{\vec s}}}
	=
	2^{-g},
\end{align}
$\vec d$ lies in the radical of $\C_{\vec s}$ and we get a constraint on $\vec s$.
If the kernel of $\mathbf{G}_{\vec d}$ is typically small, one can iterate through all candidates for $\C_{\vec s}$ and check the properties of the true $\C_{\vec s}$, namely, that $\rad \C_{\vec s}$ is doubly even and that its is given by $g$.

Since the rows of $\mat H$ are essentially random, we expect that $\mat H_{\vec d}$ has around $m/2$ rows which are linearly independent.
In the original scheme of \citet{shepherd_temporally_2009}, $n\simeq m/2$, and we thus expect $\dim \ker(\mat G_{\vec d}) \in O(1)$. 
More precisely, \citeauthor*{bremner_iqp_2023_v1} show that indeed $\mb E_{\vec d}[\dim \ker \mat G_{\vec d} ]\geq  n - m/2$. 
Thus, the runtime of the Linearity Attack scales exponentially with $n-m/2$.
In the new challenge of \citeauthor*{bremner_iqp_2023_v1}, $n=300, m=360$, so that $n-m/2 = 120$, meaning that $\ker \mat G_{\vec d} $ is so large that this simple approach is no longer feasible.

In fact, for $n - m/2 > 0 $, already the kernel of $\mat H_{\vec d}$ (which is contained in $\ker \mat G_{\vec d}$) will be nontrivial.  
But the relevant part of the kernel of $\mat G_{\vec d}$ in which the secret is hiding is independent of $n$ so long as $ g + d < n$, and only requires that $\vec d$ has zero-entries in the obfuscated coordinates of $\mat F$. 
Thus, we expect that $\dim \ker \mat G_\vec d$ is roughly independent of the event $\vec s \in \ker \mat G_{\vec d}$. 
We can thus allow ourselves to ignore large kernels in the search for $\vec s$, if we are less ambitious about exploring those very large kernels, boosting the success probability.

\begin{figure}
	\centering
	\includegraphics{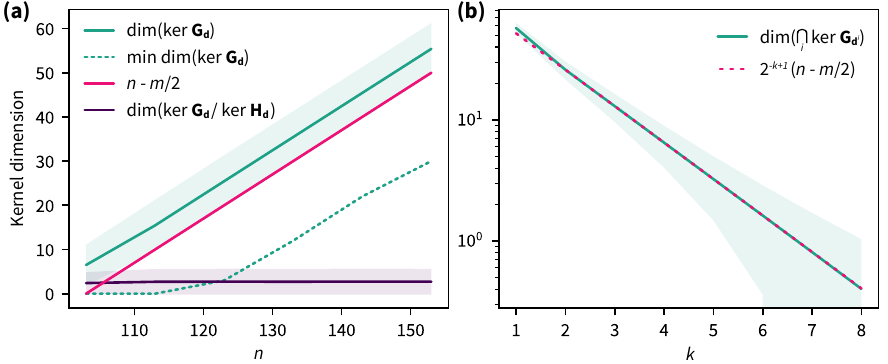}
	\caption{\label{fig:kernels}
	\emph{(a)} 
	Dimension of $\ker (\mat G_{\vec d})$ (green) 
	for the QRC construction with $q = 103$ and $m = 2q$ for 100 random instances and 1000 random choices of $\vec d$ per point.
	Shaded areas represent one standard deviation. 
	This is compared to the simplified theoretical prediction $n - m/2$ (pink).
	The dotted line designates the minimum observed value of $\dim(\ker \mat G_{\vec d})$. 
	The original Linear Attack runs in time roughly exponential in the green curve, whereas the ``lazy'' approach reduces this to about the exponential of the dotted one.
	Finally, the violet line depicts $\ker (\mat G_{\vec d}) / \ker(\mat H_{\vec d})$.
	The fact that it does not depend on $n$ is compatible with the expectation that the probability of finding the secret $\vec s$ in the kernel of any given Gram matrix $\mat G_{\vec d}$ is roughly independent of the size of the kernel. \\
	\emph{(b)} Dimension of $\bigcap_{i \in [k]}\ker (\mat G_{\vec d^i})$ (green) 
	for the QRC construction with $q = 103, n = q + r,r = (q+1)/2,m = 2q$. 
	We used 100 random instances and 1000 random choices of $\vec d^1, \ldots, \vec d^k$ per point. 
	The simple theoretical prediction of $2^{-k+1}(n - m/2)$ (dotted pink) is seen to be in good agreement with the numerical experiments up to a constant prefactor.
	}
\end{figure}

More precisely, we expect that $\langle \mat H_i, \vec d \rangle = 1$ with probability $1/2$.
Thus, the number of rows $m_{\vec d}$ of $\mat H_{\vec d}$ will follow a binomial distribution
\begin{align}\label{eqn:binom lazy}
	m_{\vec d} \sim \operatorname{Bin}\left( N \approx m, p=0.5 \right),
\end{align}
with mean $m/2$ and standard deviation $\sqrt m/2$.  
Sincemost rows of $\mat H$ are linearly independent and we expect the values of $\langle \mat H_i, \vec d \rangle$ to be only weakly correlated, we thus expect the dimension of the kernel to be given by $\dim(\ker \mat H_{\vec d} ) \approx n - m_{\vec d}$ which is roughly Gaussian around $n - m/2$ with standard deviation $\sqrt m/2$; see \cite[Theorem 3.2.2]{kolchin_random_1998} for a more precise statement.

For the Lazy Linearity Attack the relevant parameter determining the success of the attack is the probability of observing a small kernel in the tail of the distribution over kernels of $\mat H_{\vec d}$, induced by the random choice of $\vec d$.    
As discussed above, this probability decreases exponentially with $n - m/2  = g + w + i$, where we have defined the \emph{imbalance} 
\begin{align}
	i \coloneqq \frac {m_2 - m_1}2. 
\end{align}

Let the cumulative distribution function of the Gaussian distribution with mean $\mu$ and standard deviation  $\sigma$ be given by $C_{\mu,\sigma}: \mb R \rightarrow [0,1]$. 
Then, the expected endurance required for the Lazy Linearity Attack with ambition $A$ to succeed is given by 
\begin{align}
	E \sim \frac {2^g}{ C_{n - m/2, \sqrt m/2}(A)}. 
\end{align}

In numerical experiments, we find our predictions to be accurate up to a constant offset in the predicted mean of $\dim(\ker \mat H)$, see \cref{fig:kernels}a. 
In particular, the dimension of $\ker \mat G / \ker \mat H$ is independent of $n$, which is indeed evidence that there is no correlation between the size of the kernel of $\mat G$ and whether or not it contains a secret.

\subsubsection{Application to the Extended QRC construction}
\label{sec:lazy qrc}

\begin{figure}
\centering
	\includegraphics{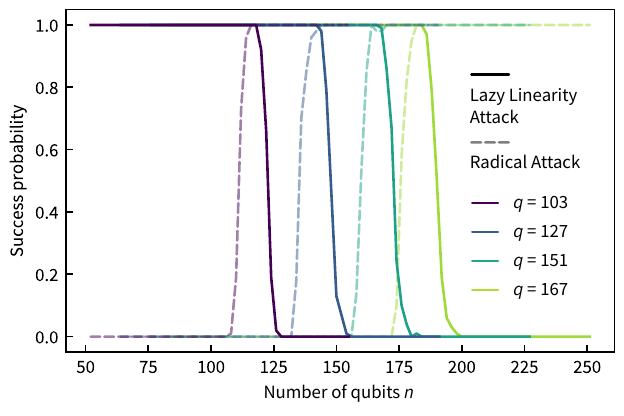}
	\caption{\label{fig:lazymeyer vs radical} 
	Performance of the Lazy Linearity Attack (solid lines) with ambition $A = 8$, endurance $E = 1000$, and significance threshold $g_{\text{th}} = 1$, and the Radical Attack (dashed lines) for values of $q = 103, 127, 151, 167$. $100$ instances per point.	
	Compared to \cref{fig:meyer vs radical}, we find that the ``lazy'' approach has extended the range of $n$ for which the Linear Attack recovers the secret with near-certainty.
	The shift is sufficient that the two algorithms now cover the entire parameter range.
}
\end{figure}

Applying the Lazy Linearity Attack to the Extended QRC construction, we find that it succeeds with near-unit success rate until $n - m/2 \sim 10$ with parameters $A = 8, E = 1000$. 
Combined with the Radical Attack, we are now able to retrieve the secrets for all proposed parameters  of the Extended QRC construction, see \cref{fig:lazymeyer vs radical}.
For the Extended QRC construction, $m_1 = m_2$, which means that $i = 0$, and hence the success probability of the Lazy Linearity Attack decreases exactly with $w  + g$ and the tunable parameters of the method.  

\subsection{The Double Meyer Attack}
\label{sec:double meyer attack}

In the previous section, we discussed how to exploit statistical fluctuations to avoid having to search through large kernels.
But as $n-m/2$ increases, this strategy will eventually fail to be effective.
We thus present another ansatz, the \emph{Double Meyer Attack}, stated as \cref{alg:doublemeyer}.
It reduces the size of the kernel, essentially by running several Linearity Attacks at once.

\begin{algorithm}
	\caption{Double Meyer Attack}\label{alg:doublemeyer}
	\begin{algorithmic}[1]
	\Function{DoubleMeyer}{$\mat H, k, g_{\text{th}}, A, E$}
	\While {$\epsilon < E $}
		\For{$i \in [k]$}
			\State Draw a uniformly random $\vec d^i \leftarrow \FF_2^n$
			\State $\mat G_{\vec d^i}\leftarrow\mat H_{\vec d^i}^T \mat H_{\vec d^i}$
			\State $\mat G \leftarrow (\mat G^T | \mat G_{\vec d^i}^T)^T $
		\EndFor
		\If {$\dim \ker \mat G < A$}
			\For{$\vec x \in \ker \mat G_{\vec d}$}
			\If{$\rad \range( \vec H_{\vec x}) \neq \{0\}$ and doubly-even and $\rank (\mat H_{\vec x}^T\mat H_{\vec x}) \le g_{\text{th}}$}
			\State \Return $\vec x$ and \textbf{exit}.
			\EndIf 
			\EndFor
			
		\EndIf	
		\State $\epsilon \leftarrow \epsilon +1$
	\EndWhile
	\State \Return ``fail''
	\EndFunction
	\end{algorithmic}
\end{algorithm}

\subsubsection{Analysis}

At a high-level, the Double Meyer examines the elements of the intersection of the kernels of several $\mat G_{\vec d^i}$, where the $\vec d^i$ are uniformly distributed random vectors \footnote{
	The essential generalization over Kahanamoku-Meyer's Linearity Attack can be gleamed from the simplest special case $k=2$, which, of course, explains the name we have adopted for this ansatz.
}.
Since the events $\mat H_{\vec s } \vec d^i \in \rad \C_{\vec s}$ are independent for different $i$'s, we have that
\begin{align}
	\Pr\left [\vec s  \in \bigcap_{i \in [k]} \ker \mat G_{\vec d^i} \right] = (2^{-g})^k = 2^{-gk}. 
\end{align}
At the same time, we expect $\dim (\bigcap_{i \in [k]} \ker \mat G_{\vec d^i}) \approx 2^{-k+1} ( n - m/2)$. 
To see this, observe that the kernel of $\mat G_{\vec d^i}$ decomposes as  $\ker \mat G_{\vec d^i} = \ker \mat H_{\vec d^i} + \rad \range(\mat H_{\vec d^i})$. 
For random vectors $\vec d^i$ we then expect $\rad \range(\mat H_{\vec d^i})$ to be iid.\ random subspaces, while the kernels of $\mat H_{\vec d^i}$ are correlated, since every pair of $\ker \mat H_{\vec d^1}, \ker \mat H_{\vec d^2}$ shares half the rows. 
Thus, the intersection of $\rad \range \mat H_{\vec d^i}$ decays exponentially with $k$, while the intersection $\bigcap_{i \in [k]}\ker \mat H_{\vec d^i}$ decreases much faster as $n - (1- 2^{-k}) m$.  
Altogether, the intersection decomposes into sums of intersections (for $k=2$)
\begin{align}
	\bigcap_{i=1}^2 \ker \mat G_{\vec d^i} = (\ker \mat H_{\vec d^1} + \rad \range(\mat H_{\vec d^1})) \cap (\ker \mat H_{\vec d^2} + \rad \range(\mat H_{\vec d^2})),
\end{align}
where the exponential decay as $2^{-k}$ stemming from the last term dominates the scaling. 
%

Choosing $k \sim \log n$ is sufficient to reduce the kernel dimension to $O(1)$. 
Moreover, $g$ needs to be of order $O(\log n)$ in order to maintain a sample-efficient verification test for the challenge. 
Thus,
$2^{gk} \in 2^{O(\log^2 (n))}$,
i.e.\ the Double Meyer Attack is expected to run in at most most quasi-polynomial time, for any choice of $n,m$. 
However, even moderate values of $g$ will make this approach infeasible in practice. 
For the challenge data set, we expect that good parameter choices are $k = 6$, $A = 3$ and $E\gtrsim 2^{gk} = 2^{24} \approx 10^{7.22}$, though we have not spent enough computational resources to have recovered a secret in this regime.

We observe that slack between the threshold rank $g_{\text{th}}$ and the true $g$ often leads to a misidentified secret. 
This is explained by vectors $\vec v \in \bigcap_{i \in [k]}\ker \mat H_{\vec d^i}$ whose image under $\mat H$ has low Hamming weight $|\mat H \vec v| \leq g_{\text{th}} - g$.
This corresponds to rows of $\mat H$ which can be mapped to $s = g_{\text{th}} - g$ unit vectors $\vec e^1, \ldots, \vec e^s$ which are linearly independent of the first $m_1 $ rows of $\mat H$. 
These vectors may be absorbed into $\mat F$, adding $s$ nontrivial columns to it, while adding a zero row to $\mat D$, which keeps its range doubly even. 
The alternative secrets $\vec v$ found in this way are also observable in the sense that they satisfy $\Pr_{\vec x \leftarrow D_{\mat H}} [\langle \vec x, \vec v\rangle = 0 ] = (2^{- (g + s)/2} +1)/2$. 
In order to find the ``true'' secret, one should therefore run the attack for increasing values of $g_{\text{th}}$, and halt as soon as a valid secret is found. 

The low-Hamming weight vectors identified above inform the final ansatz of this paper, \emph{Hamming's Razor}, presented in \cref{sec:hammings razor}.

\subsubsection{Application to the Extended QRC construction}

We find that the Double Meyer Attack recovers the secrets of the Extended QRC construction with near-certainty in all parameter regimes proposed by \citeauthor*{bremner_iqp_2023_v1}.
For the QRC construction, $g = 1$ and hence the runtime of the Double Meyer Attack is just given by $2^k$. 
Choosing $k=6, g_{\text{th}}= 1, E = 8, A = 1000$ is sufficient to recover the secret in all of $100$ random instances of the Extended QRC construction for all values of $n \in [r, q+r]$ and $q \in \{103,127,151,167\}$.

\subsubsection{Further improvements}
\label{sec:round2}

As stated, the Double Meyer draws vectors $\vec d^i$ uniformly at random, in the hope that 
$\mat H_{\vec s} \vec d^i \in \rad \C_{\vec s}$.
But there might be more efficient ways of obtaining vectors $\vec d^i$ satisfying this condition.
For instance, under the assumptions of Corollary~\ref{cor:kernel}.1, \emph{any} element $\vec d$ of 
$\mat H(\ker( \mat H^T \mat H))$ 
has this property.
Adding such vectors to the collection of $\vec d^i$'s therefore provides additional constraints for $\vec s$ at essentially no computational cost.

In particular, this modification of the Double Meyer would break a variant of the construction \citet{private_communication} proposed as an initial reaction to the pre-print of this paper 
with parameters \texttt{--AB-type zero --concat\_D --concat\_C1} as implemented in commit \texttt{7d3bd3} of their github repository \cite{cheng_repo_update}.

\subsection{Hamming's Razor}
\label{sec:hammings razor}

In this section, we describe a method that allows one to ``shave off'' certain rows and columns from $\mat H$ without affecting the code space $\C_{\mat s}$.
Such redundancies can be identified given a vector $\vec d \in \FF_2^n$ such that $\mat H \vec d$ has low Hamming weight.
The method comes in two varieties:
The simpler \emph{Singleton Razor}, discussed first, and the more general \emph{Hamming's Razor} proper.

\subsubsection{The Singleton Razor}

Let us agree to call $i\in[m]$ a \emph{singleton for $\mat H$} if there is a solution to $\mat H \vec d = \vec e^i$.

We discuss the idea based on the unobfuscated picture of \cref{eqn:blockmat}.
Assume that $\supp \range \mat D=[m_1]$. 
Then there is no singleton among the first $m_1$ coordinates, because $\vec e^i$ is not orthogonal to $\range \mat D$, while all vectors in the range of $(\mat F \,|\, \mat D)$ are.
Thus, knowing a singleton $i$, one may trim away the $i$-th row of $\mat H$ without affecting the code space.
Alternatively, one can perform a coordinate change on $\FF_2^n$ that maps the corresponding pre-image $\vec d$ to $\vec e^n$ and then drop the $n$-th column of $\mat H$.
In fact, both operations may be combined, without changing $\C_{\mat s}$.

The generator matrix $\mat H$ of the challenge data set affords $69$ singletons.
All singletons do indeed belong to redundant rows \cite{repository}. 
The second part of \cref{lem:hamming} below suggests an explanation for this surprisingly high number.

\subsubsection{Hamming's Razor}

We now generalize the singleton idea to higher Hamming weights.
Starting point is the observation that 
$\range \mat C$ 
can be expected to contain vectors of much lower Hamming weight than
$\range \mat (\mat F\,|\,\mat D)$.
This will lead to a computationally efficient means for separating redundant from non-redundant rows.

The following lemma collects two technical preparations.

\begin{lemma}
	\label{lem:hamming}
	Let $\mat M$ be an $m\times (m-h)$ binary matrix chosen uniformly at random.
	The probability that the minimal Hamming weight of any non-zero vector in the range of $\mat M$ is smaller than $k$ is 
	exponentially small in $k_1 - k$, where
	\begin{align*}
		k_1 = h \frac{\ln 2}{\ln m + 2} \approx \frac{h}{\log_2 m}.
	\end{align*}
	More precisely and more strongly, the probability is no larger than
	$e^{-\lambda (k_\infty-k)}$, for
	\begin{align}\label{eqn:hamming_sequence}
		k_\infty = \lim_{i\to\infty} k_i,
		\qquad
		k_i
		=k_1
			+ 
			\frac{
				(k_{i-1}-1) \ln k_{i-1}
		}{\ln m+2},
		\qquad
		\lambda  = \frac{1}{k_\infty} + \ln\frac{m}{k_\infty} > 0.
	\end{align}

	Conversely, let $\mat M$ be any binary matrix whose range has co-dimension $h$.
	For $S\subset[m]$, let $V_S\subset \FF_2^m$ be the subspace of vectors supported on $S$.
	Then $\range M$ has a non-trivial intersection with $V_S$ if $|S|>h$.
\end{lemma}

\begin{proof}
	Let $\vec v$ be a random vector distributed uniformly in $\FF_2^{m}$.
	Then, as long as $k\leq m/2$,
	\begin{align*}
		\Pr\left[|\vec v| \leq k\right]
		=
		\sum_{k'\leq k}
		\binom{m}{k'} 2^{-m}
		\leq
		k
		\binom{m}{k} 2^{-m}
		\quad
		\Rightarrow
		\quad
		\Pr\left[\min_{0\neq \vec v \in \range \mat M} |\vec v| \leq k\right]
		\leq
		k
		\binom{m}{k} 2^{-h}.  
	\end{align*}
	Using the standard estimate
	$ \ln \binom{m}{k} \leq k (\ln m+1) - k \ln k $,
	the logarithm of the bound is
	\begin{align*}
		l(k):=
		k (\ln m+1) - (k-1) \ln k - h \ln 2 .
	\end{align*}
	The function $l(k)$ is concave, negative at $k=0$, positive at $k=m/2$, and thus has a zero in the interval $[0,m/2]$.
	What is more, $l(k)=0$ if and only if
	\begin{align*}
		k 
		=
		h
		\frac{
			\ln 2 
		}{
		\ln m+1
		}
			+ 
			\frac{
			(k-1) \ln k
		}{\ln m+1} .
	\end{align*}
	Because the r.h.s.\ is monotonous in $k$, the recursive formula (\ref{eqn:hamming_sequence}) defines an  increasing sequence $k_i$ of lower bounds to the first zero.
	As a non-decreasing sequence on a bounded set, the limit point $k_\infty$ is well-defined.
	Due to concavity,
	$l(k)$ is upper-bounded by its first-order Taylor approximations.
	The claim then follows by expanding around $k=k_\infty$.

	The converse statement is a consequence of the standard estimate
	\begin{align*}
		\dim \big( \range \mat M \cap V_S) \geq \dim \range \mat M  + \dim V_S - m.
	\end{align*}
\end{proof}

Choose a set $S\subset[m]$ and let $\mat H_{\setminus S}$ be the matrix obtained by deleting all rows of $\mat H$ whose index appears in $S$.
Let $S_1 = S \cap [m_1]$ be the intersection of $S$ with the ``secret rows''  
and $S_2 = S \cap [m_1+1, m_2]$ the intersection with the redundant rows.
Then $\vec d\in\ker\mat H_{\setminus S}$ if and only if 
\begin{align*}
	\supp (\mat H_{\vec s} \vec d) \subset S_1
	\quad\text{and}\quad
	\supp (\mat R_{\vec s} \vec d) \subset S_2.
\end{align*}
Now model $(\mat F\,|\,\mat D)$ as a uniformly random matrix.
\cref{lem:hamming} applies with $m=m_1, h=m_1-g-d$, giving rise to an associated value of $k_\infty$.
If $|S_1| < k_\infty$ then, up to an exponentially small probability of failure, the first condition can be satisfied only if $\mat H_{\vec s} \vec d = 0$.
Therefore, each non-zero element $\vec d \in \ker \mat H_{\setminus S}$ identifies $\supp(\mat H \vec d)$ as a set of redundant rows, which can be eliminated as argued in the context of the Singleton Attack.

This observation is useful only if it is easily possible to identify suitable sets $S$ and non-zero vectors $\vec d$ in the associated kernel.
Here, the second part of \cref{lem:hamming} comes into play.
If $|S_2|>m_2 - (n-g-d)$, then by the lemma and \cref{eqn:c-has-full-rank}, there exists a non-zero $\vec d$ such that $\supp(\mat C \vec d)\subset S_2$.

This suggests to construct $S$ by including each coordinate $i\in[m]$ with probability $p$, chosen such that 
$p m_2 > m_2 - (n-g-d)$
and 
$p m_1 < k_\infty$.
For the challenge parameters, these requirements are compatible with the range $[0.01, 0.13]$ for $p$.

In fact, one can base a full secret extraction method on this idea, see \cref{alg:hammingrazor}. 
Repeating the procedure for a few dozen random $S$ turns out to reveal the entire redundant row set, and thus the secret, for the challenge data \cite{repository}.
The attack may be sped up by realizing that the condition on $|S_1|$ was chosen conservatively.
The first part of \cref{lem:hamming} states that the \emph{smallest} Hamming weight that occurs in the range of $\mat H_{\vec s}$ is about $k_\infty$.
But a \emph{randomly chosen} $S_1$ of size larger than $k_\infty$ is unlikely to be the support of a vector in $\range \mat H_{\vec s}$ unless $|S_1|$ gets close to the much larger second bound in the lemma.
This optimization, for a heuristically chosen value of $p=0.25$, is used in the sample implementation provided with this paper \cite{repository} and recovers the secret with high probability.

\begin{algorithm}
	\caption{Hamming's Razor}\label{alg:hammingrazor}
	\begin{algorithmic}[1]
	\Function{HammingRazor}{$\mat H, p, E$}
	\State $S \leftarrow \emptyset$.
	\For {$\_\in  [E] $}
		\State Draw a random vector $\vec d \in \FF_2^m$ with entries $\vec d_i \leftarrow \operatorname{Bin}(\FF_2,p)$. 
		\State $\mat H[\vec d] \leftarrow \operatorname{diag}(\vec d) \mat H$. 

		\State $\mat K \leftarrow$ a column-generating matrix for $\ker \mat H[\vec d]$.
		\State Append the support of the columns of $\mat H \mat K$ to  $S$. 
	\EndFor
	\State Solve the $\FF_2$-linear system $\mat H \vec s = \vec 1_{S^c}$
	\State \Return $\vec s$
	\EndFunction
	\end{algorithmic}
\end{algorithm}



\section{Discussion}
\label{sec:conclusion}

In this work, we have exhibited a number of approaches that can be used to recover secrets hidden in obfuscated IQP circuits.

As a reaction to a draft version of this paper, Bremner, Cheng, and Ji have modified their proposal to evade the attacks described here.
A first update (communicated privately) led to our improved Double Meyer, as sketched in Sec.~\ref{sec:round2}.
As of September 2024, the authors have provided us with a version of their protocol in which we have not found a weakness~\cite{bremner_iqp_2023_v2}.

It may be instructive to compare the situation to the more mature field of classical cryptography, which benefits from a large public record of cryptographic constructions and their cryptanalysis.
New protocols can thus be designed to resist known exploits and be vetted against them. 
In particular, most cryptographic protocols actually in use are not rigorously proven to be secure.
Instead, trust in them is based on a long and public history of constructions, as well as successful and unsuccessful attempts at attacking them.
The well-documented story of \emph{differential cryptanalysis} provides an instructive example.

In this light, we consider the high-level contributions of our paper to be this:
1.\ It shows potential users of crytpographically backed-up demonstrations of quantum supremacy that previous proposals have been broken repeatedly, so that their security should not be taken for granted. 
2.\ The fact that we have not yet been able to identify an efficient attack against the latest version of the protocol should raise one's trust in that version of the proposal, compared to a protocol that has not been the subject of a public security review.
3.\ Our attacks clarify properties of the IQP-based protocol that make it amenable to classical attacks as well as a collection of cryptanalyltic techniques exploiting those. These tools can guide and must be taken into account by designers of future constructions.

At the same time, our cryptanalysis remains to have two important consequences on IQP-based verified quantum advantage using the construction of \citet{bremner_iqp_2023_v1}.
First, our \emph{Double Meyer} attack remains valid for all instances and has quasi-polynomial runtime $2^{-\Omega(\log^2 n)}$ given the verification condition that the signal $2^{-g}$ remains inverse polynomially large.  This is a significant improvement over the previous state of the art, which was an exponential-time algorithm.
Second, in order to circumvent our attacks, \citet{bremner_iqp_2023_v2} have had to significantly increase the problem or `key' size from the original challenge data with parameters $n=300, m=360, g=4$ \cite{} to $n=700$ qubits, $m = 1200$ Hamiltonian terms and a verification signal of $2^{-g} = 2^{-10}$. 
It is therefore an interesting open problem to compare the implementation cost of this scheme to other resource-efficient schemes which come with provable security guarantees based on well-studied classical assumptions \cite{kahanamoku-meyer_classically_2022,zhu_interactive_2023}. 

We emphasize that, in our judgment, the problem of finding ways to efficiently certify the operation of near-term quantum computing devices is an important one and the idea to use obfuscated quantum circuits remains appealing.
More generally, the story of IQP-based verified quantum advantage and our contribution to it illustrate that results that exhibit weaknesses in published constructions should not cause the community to turn away, but should rather serve as sign posts guiding the way to more resilient schemes.
We remain curious whether the security of the new construction of \citet{bremner_iqp_2023_v1} holds up to further scrutiny by the community. 

%

\section*{Acknowledgements}

First and foremost, we are grateful to Michael Bremner, Bin Cheng, and Zhengfeng Ji for setting the challenge and providing a thorough code base that set a low threshold for its study. 
We feel it is worth stating that we have tried numerous times to construct alternative hiding schemes ourselves---but found each of our own methods much easier to break than theirs.
We also thank Mick, Bin, and Zhengfeng for detailed comments and discussions on earlier versions of this paper.

We are grateful for many inspiring discussions on hiding secrets in a circuit description, with Dolev Bluvstein, Alex Grilo, Michael Gullans, Jonas Helsen, Marcin Kalinowski, Misha Lukin, Robbie King, Yi-Kai Liu, Alex Poremba, and Joseph Slote.

DG's contribution has been supported by 
the German Ministry for Education and Research under the QSolid project, 
and by
Germany's Excellence Strategy -- Cluster of Excellence Matter and Light for Quantum Computing (ML4Q) EXC 2004/1 (390534769).
DH acknowledges funding from the US Department of Defense through a QuICS Hartree fellowship and from the Simons Institute for the Theory of Computing, supported by DOE QSA.

\bibliographystyle{apsrev4-1}

\bibliography{library,shortened_lib}

\begin{appendix}

\section{Implementation details}
\label{sec:implementation-details}

While running the software package provided with \cite{bremner_iqp_2023_v1} tens of thousands of times, we found a number of extremely rare edge cases that were not explicitly handled.

In particular, the \verb|sample_parameters()| and the \verb|sample_D()| functions would very rarely return inconsistent results.
Rather straight-forward corrections are published in \cite{repository}.

Another possible discrepancy between the procedure described in the main text of Ref.~\cite{bremner_iqp_2023_v1} and their software implementation concerns the generation of the ``redundant rows''.
The issue is a little more subtle than the first two, so we briefly comment on it here.

In the paper, the relevant quote is
\begin{quote}
	``Therefore, up to row permutations, the first $n - r$ rows of $\mat R_{\vec s}$ are sampled to be 
	\emph{random independent} rows that are orthogonal to $\vec s$ and lie outside the row space of $\mat H_{\vec s}$.''
\end{quote}
(emphasis ours).
The computer implementation is given by the
\verb|add_row_redundancy()| function in \verb|lib/construction.py|,
in particular by these lines:
\begin{verbatim}
s_null_space = s.reshape((1, -1)).null_space()

full_basis = row_space_H_s
for p in s_null_space:
  if not check_element(full_basis.T, p):
  full_basis = np.concatenate((full_basis, p.reshape(1, -1)), axis=0)

R_s = full_basis[r:] # guarantee that rank(H) = n
\end{verbatim}
At this point, the first $n-r$ rows of $R_s$ are not  ``random independent''.
The behavior of this piece of code depends on the detailed implementation of the \path{null_space()} function, which we do not directly control.
While the obfuscation process will later add randomness,
we caution that the \emph{same} random invertible matrix $\mat Q$ acts both on $\mat R_{\vec s}$ and on $\mat H_{\vec s}$.
Any relation between these two blocks that is invariant under right-multiplication by an invertible matrix will therefore be preserved.
As a mitigation of this possible effect, we suggest to add an explicit randomization step, such as
\begin{verbatim}
	s_null_space=rand_inv_mat(s_null_space.shape[0],seed=rng)@s_null_space
\end{verbatim}
to the routine.
(Though in practice, we did not observe different behavior between these two versions).

The code published in \cite{repository} contains three implementations of the \verb|add_row_redundancy()| function.
The version used to create the challenge data (c.f.\ Sec.~\ref{sec:challenge}),
the one published with Ref.~\cite{bremner_iqp_2023_v1},
and finally the one with the explicit extra randomization step added.
The numerical results reported in the main part of this paper were generated by the third routine, though we have also include 20k runs performed with the second version \cite{repository}.
The effectiveness of the Radical Attack does not seem to differ appreciatively between these two implementations.

\end{appendix}

\end{document}